\documentclass[epj,final]{svjour}

\usepackage{graphicx}
\usepackage{amsmath}
\usepackage{amssymb}
\usepackage{amscd}

\begin{document} 
 
\title{Stability analysis of a noise-induced Hopf bifurcation}
\author{Kirone Mallick\inst{1} \and Philippe Marcq\inst{2}}                 

\institute{Service de Physique Th\'eorique, Centre d'\'Etudes de Saclay,
 91191 Gif-sur-Yvette Cedex, France \\
\email{mallick@spht.saclay.cea.fr}\and 
Institut de Recherche sur les Ph\'enom\`enes Hors \'Equilibre,
Universit\'e de Provence,\\
49 rue Joliot-Curie, BP 146, 13384 Marseille Cedex 13, France\\
 \email{marcq@irphe.univ-mrs.fr}}

\date{August 8, 2003}
\abstract{
We study analytically and numerically  the  noise-induced transition 
between an absorbing and an oscillatory state in a Duffing oscillator 
subject to multiplicative, Gaussian white noise. We show in a non-perturbative
 manner that a stochastic   bifurcation  occurs when  the 
Lyapunov exponent of the linearised system becomes positive.
 We deduce from  a simple formula  for the Lyapunov exponent  
  the phase diagram of the stochastic Duffing oscillator.
The behaviour of  physical observables, such as the oscillator's mean energy, 
is studied both close to and far from  the bifurcation.
} 

\maketitle 

\section{Introduction}
\label{sec:intro}

     The stability   of a dynamical system can be strongly   
affected  by the presence of uncontrolled
 perturbations such as  random noise 
 \cite{lefever}.  Noise-induced bifurcations  are of primary
  interest in many fields of applied mathematics and physics and   occur 
 in various  experimental contexts,  as  for instance:
     electronic circuits  \cite{strato},
    mechanical oscillators \cite{landaMc},
   laser physics  \cite{lefever},
 surface waves \cite{fauve}, thermal and electrohydrodynamic
 convection in fluids  or liquid crystals \cite{kramer1,kramer2,behn} and 
 diffusion in  random media \cite{frisch}. In nonlinear dynamical systems,
 the  interplay of noise and  nonlinearity
  may produce  unusual  phenomena: 
 noise can shift   the  bifurcation  threshold 
 from one phase to another \cite{luecke},
 it can  induce new  phase-transitions in spatially-extended systems
\cite{vandenb1}, or create spatial patterns 
\cite{vandenb2} (for a recent review see \cite{toral}).

 One of the simplest systems that may serve  as a pa\-ra\-digm  for
 noise-induced transitions is the  nonlinear oscillator with
 parametric noise \cite{lindenberg,landa,drolet}, i.e., 
with a frequency that
 fluctuates randomly with time around  a given  mean value
 \cite{bourret,bourretFrisch}.  
 In a recent work \cite{philkir1,philkir2}, we studied  the time
  asymptotic behaviour  of such an oscillator in the small damping limit. 
 We showed  that  physical observables grow  algebraically  with time
  until the dissipative time scale is reached;
 then  the system  settles  into an oscillatory 
 state described by a stationary probability distribution function (PDF)
 of the energy $E$: the zero equilibrium  point  ($E = 0$) 
 is unstable in the small damping limit. In the present work,
 we investigate the stability of the origin for arbitrary 
 damping rate and noise strength.

 For  random   dynamical systems,  it   was recognized early on that
  various `naive'  stability criteria \cite{bourret,bourretFrisch}, 
obtained by linearizing the  
dynamical equation  around the  origin, lead to ambiguous results.  
(This is in  contrast with  the  deterministic  case    for which
 the bifurcation threshold  is  obtained  without ambiguity
 by studying the eigenstates of  the linearized  equations
  \cite{manneville}.) 
  In fact,   it was conjectured in \cite{bourretFrisch}
 and proved in  \cite{lindenberg3} that,   in a linear oscillator 
with arbitrarily small parametric noise, all moments beyond a certain 
order diverge in the long time limit.  Thus,  any   criterion based on 
 finite mean  displacement,  momentum,  or energy of the  linearized dynamical
 equation   is not  adequate   to insure global stability: the bifurcation 
threshold  of a  nonlinear random dynamical system cannot be determined  
simply   from  the moments of the linearized system. 
 In practice, the transition point is usually calculated 
 in a perturbative manner by using weak noise expansions
 \cite{landaMc,luecke,landa,drolet}.

In the present  work, we shall obtain in a non-per\-tur\-ba\-tive manner
the bifurcation threshold of a stochastic Duffing oscillator 
with multiplicative, Gaussian white noise. 
We use a technique described in \cite{philkir1} where
the stochastic dynamical equations are expressed
in terms of energy-angle variables. 
Thanks to a detailed analysis of the associated Fokker-Planck equation,
we shall  show  that the bifurcation occurs precisely
where  the Lyapunov exponent of the \emph{linear} oscillator changes sign.
This  finding is confirmed by numerical simulations. The stability of 
the fixed point $E=0$ of the  (nonlinear)  Duffing oscillator 
with parametric noise  is  deduced, in this sense, 
from the linearized  dynamics.
We shall also derive the scaling behaviour of physical 
observables in the vicinity of the
bifurcation with respect to the distance to threshold.
Finally, we study the behaviour of observables far 
from the bifurcation, in the small damping or strong noise limit.

    A rigorous mathematical theory of random dynamical systems
 has been recently    developed  and  theorems relating 
 the stability of the solution of a stochastic differential
 equation to the sign of Lyapunov exponents have
 been proved \cite{arnold}.  In the present work,  we do not
 rely upon the sophisticated tools of the general mathematical
 theory but follow an intuitive approach, easily accessible
 to physicists, based upon a factorisation hypothesis of the
 stationary PDF in the limit of small and large energies. 

 This article is organized  as follows. In Section~\ref{sec:Duffing},
 we  define the system considered, present the relevant phenomenology,
and introduce  energy-angle variables for the stochastic
Duffing oscillator.
  We   study the stationary PDF of the energy in Section~\ref{sec:pdf}.
In Section~\ref{sec:lyapunov}, we derive the stability criterion for 
the stochastic Duffing oscillator and obtain its phase diagram.
 In Section~\ref{sec:scaling}, we  perform a local analysis  
 of the  physical observables in the vicinity of and far from
 the bifurcation threshold.  Section~\ref{sec:disc} is devoted
 to some concluding remarks. In Appendix~\ref{sec:identity}, we  prove some  useful 
 identities  satisfied by the  Lyapunov exponent.
 In Appendix~\ref{sec:pdfangle}, an explicit formula for 
 the  Lyapunov exponent is derived.

\begin{figure}
\includegraphics*[width=0.75\columnwidth]{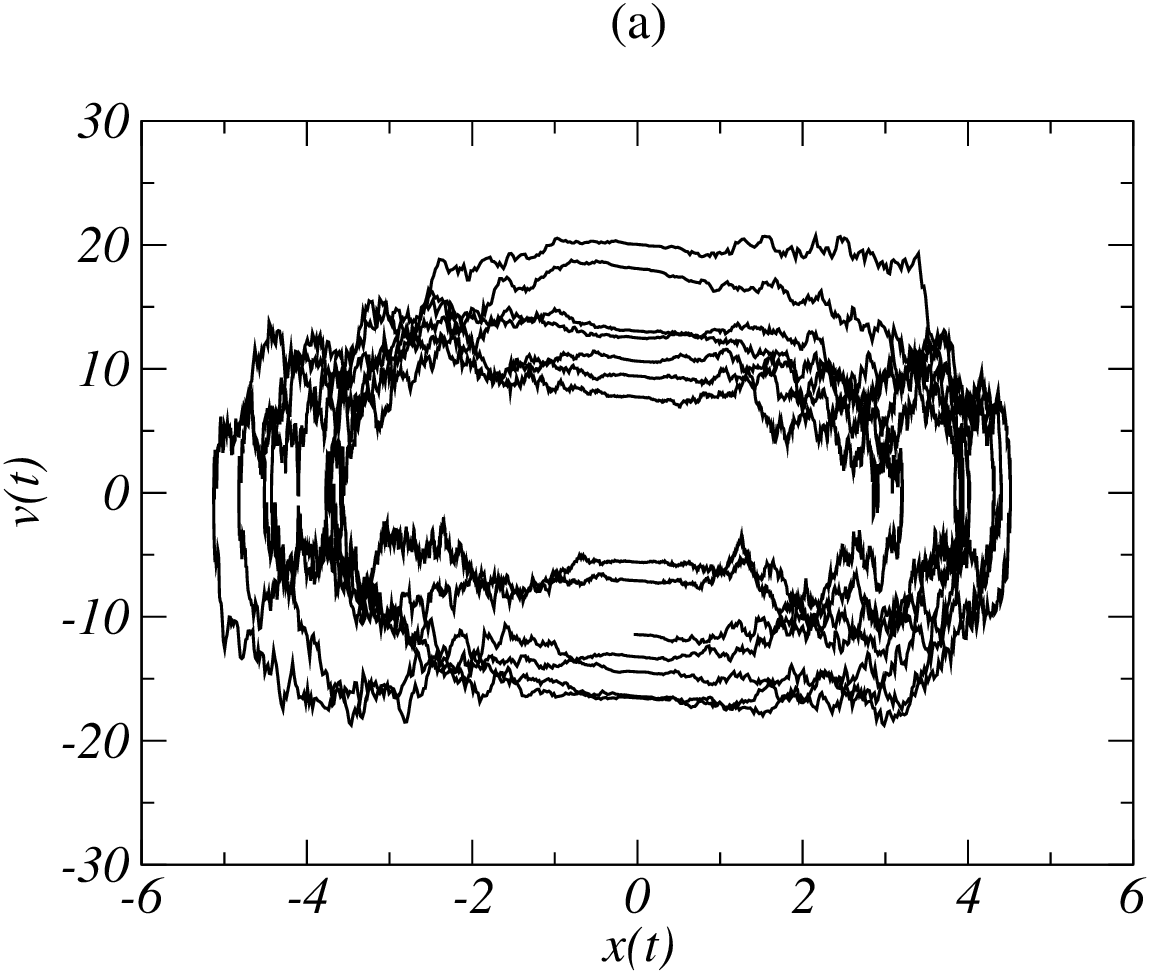}
\bigskip

\includegraphics*[width=0.75\columnwidth]{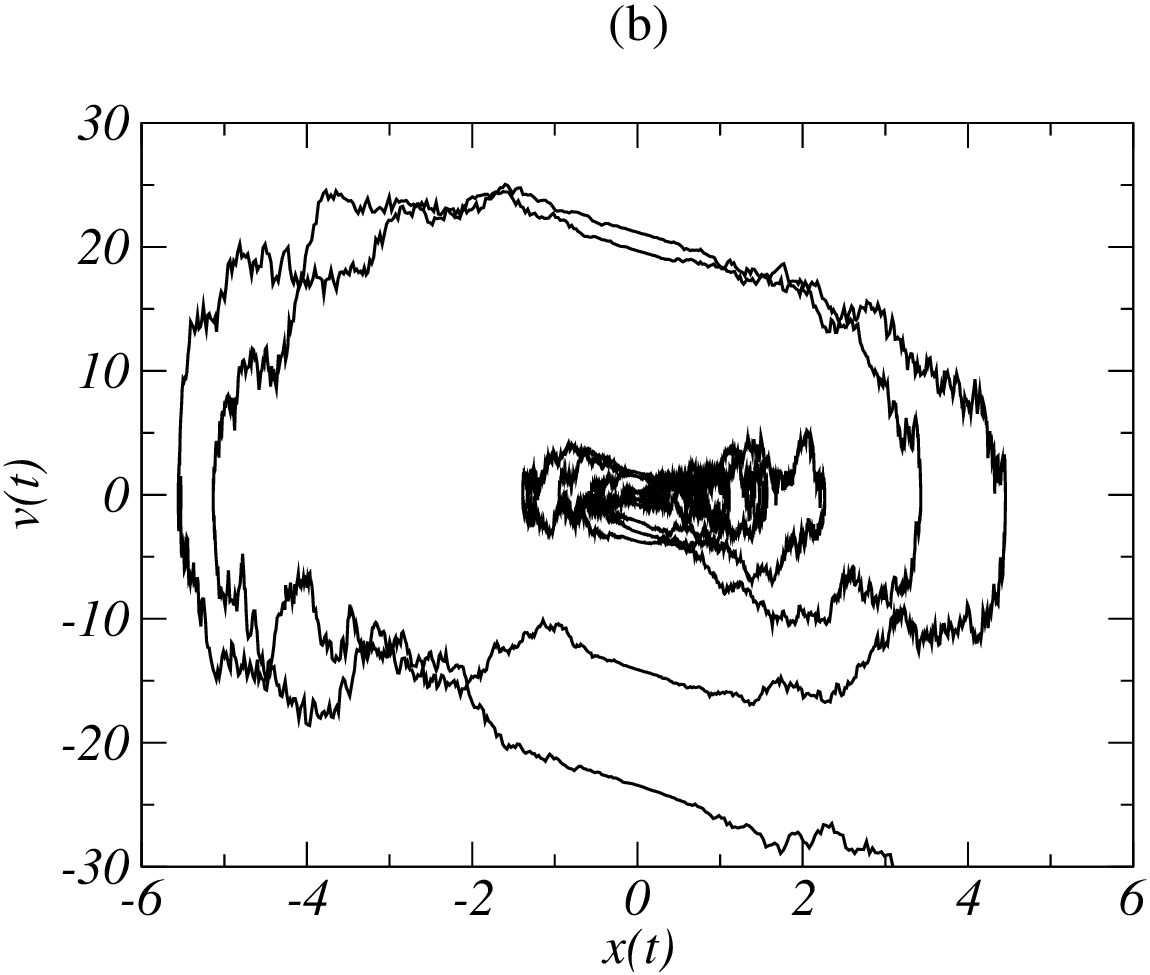}
\bigskip

\includegraphics*[width=0.75\columnwidth]{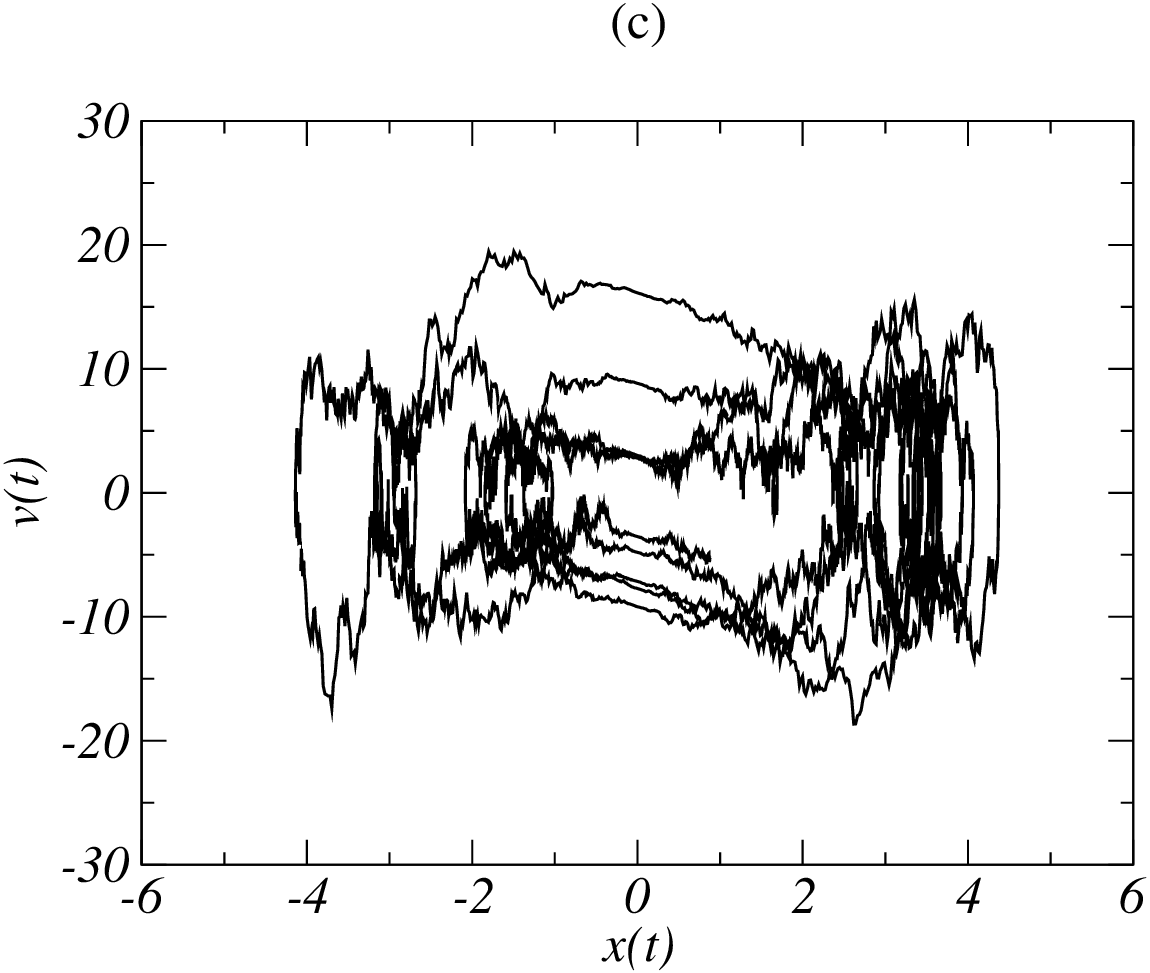}

\caption{Phase plane plots of a typical trajectory of the noisy Duffing
oscillator. Eq.~(\ref{dissipDuff}) is integrated numerically
with a time step $\delta t = 5 \ 10^{-4}$. (a) $\alpha = 0.5$,
$\Delta = 20$, $10 \le t \le 100$; (b) $\alpha = 2.0 $,
$\Delta = 20$, $t \le 100$; (c) $\alpha = 2.0$,
$\Delta = 60$, $10 \le t \le 100$. Early-time relaxation
towards noisy oscillations has been omitted for clarity in cases
(a) and (c). The trajectory spirals down towards the globally stable origin 
in case (b).}
\label{fig:traj} 
\end{figure}

\section{The Duffing oscillator with multiplicative white noise}
\label{sec:Duffing}

\subsection{Notations and phenomenology}
\label{sec:Duffing:phen}

 A nonlinear oscillator with a  randomly varying  frequency due  to 
  external noise can  be  described   
by the following equation 
\begin{equation}
\frac{\mathrm{d}^2 x}{\mathrm{d} t^2} + 
  \gamma \frac{\mathrm{d} x}{\mathrm{d} t}
  + ( \omega^2 + \xi(t)) x +
 \frac { \partial{\mathcal U}(x)}{\partial x}   = 0 \, ,
 \label{eqsto}
\end{equation}
where $x(t)$ is the position of the oscillator at time $t$,
${\mathcal U}(x)$   an anharmonic confining potential and  $\gamma$
 the (positive) friction coefficient.
 The  linear frequency of the oscillator has a mean value 
 $\omega$  and its  fluctuations  are modeled   by 
 a  Gaussian white noise  $\xi(t)$ of zero mean-value
 and of amplitude  ${\mathcal D}$ 
 \begin{eqnarray}
       \langle \xi(t)  \rangle &=&   0   \, ,\nonumber \\
 \langle \xi(t) \xi (t') \rangle &=& {\mathcal D} \, \delta(t - t')   \, .  
 \label{statxi}
\end{eqnarray} 
In this work, all stochastic differential equations are interpreted
according to Stratonovich calculus.  We shall  study the   Duffing oscillator
 with multiplicative noise,    the confining
 potential ${\mathcal U}$  being given by 
\begin{equation}
{\mathcal U}(x) = \lambda \frac{x^4}{4} \,.
\end{equation} 
 By  rewriting  time  and amplitude in dimensionless units,
  $ t := \omega t$  and   $ x:= \lambda^{1/2} \omega^{-1} x,$
    respectively, 
  Eq.~(\ref{eqsto})  becomes 
\begin{equation}
\frac{\mathrm{d}^2 x}{\mathrm{d} t^2} + 
\alpha  \, \frac{\mathrm{d} x}{\mathrm{d} t} + x + x^3 = x \, \Xi(t) ,
 \label{dissipDuff}
\end{equation}  
where  $\Xi(t)$ is a delta-correlated Gaussian  variable:
  \begin{eqnarray}
       \langle \Xi(t)  \rangle &=&   0   \, ,\nonumber \\
 \langle \Xi(t) \Xi (t') \rangle &=& \Delta \, \delta(t - t')   \, .  
 \label{defXi}
\end{eqnarray} 
 The parameters 
\begin{equation}
  \alpha = \gamma/\omega \,\,\,\,\, \hbox{ and  }
      \,\,\,\,\,     \Delta = {\mathcal D}/\omega^3  \label{parametre}
\end{equation} 
correspond to dimensionless dissipation rate and to noise strength, 
respectively. In the rest of this work, all physical quantities
are expressed in dimensionless units.

In Fig.~\ref{fig:traj}, we present, for three values of the parameters,
a trajectory in the phase plane $(x, v = \dot x)$ characteristic
of the oscillator's behaviour. Here as well as for other
simulations presented in this article, Eq.~(\ref{dissipDuff}) is 
integrated numerically by using the one step collocation method advocated in 
\cite{mannella} and described in detail in \cite{philkir1}. 
Initial conditions are chosen far from the origin, with an 
amplitude of order $1$. Noisy oscillations are observed
for small values of the damping parameter $\alpha$
[See Fig.~\ref{fig:traj}.a]. A larger value of $\alpha$ turns
the origin into a global attractor for the dynamics 
[See Fig.~\ref{fig:traj}.b].
Increasing the noise amplitude $\Delta$ at constant $\alpha$
makes the origin unstable again [See Fig.~\ref{fig:traj}.c].

For all values of the parameters $(\alpha, \Delta)$,
the origin  $(x = \dot{ x} = 0)$  is a  fixed point of
 the random dynamical system~(\ref{dissipDuff}). 
 Our main  goal, in this work, is to investigate
  the stability properties  of  the fixed point at the origin 
when the control parameters vary.

\subsection{Energy-angle representation of the  Duffing oscillator}
\label{sec:Duffing:energyangle}

 We  now  follow the method presented
 in \cite{philkir1} and introduce the  energy-angle variables
 associated  with  the Duffing oscillator.
 The deterministic and undamped
 oscillator corresponding to Eq.~(\ref{dissipDuff})
 is an integrable dynamical system  because the mechanical energy $E$,
 defined as  
\begin{equation}
   E =  \frac{1}{2}\dot x^2 + \frac{1}{2} x^2 +
   \frac{1}{4} x^4 \, ,
 \label{DuffingE}
 \end{equation}
 is a conserved quantity. Introducing
 the  angular variable $\phi$, the constant
 energy  orbits of the Hamiltonian system 
  ${\ddot x} + \, x  +  \, x^3  = 0$   are parametrized as
  \begin{eqnarray}
          x  &=& \left(  \frac{ 4E^2}{ 4E + 1}\right) ^{1/4}  
 {\rm sd}\left( \phi \;  ;k  \right),
  \label{dufx}  \\                 
     \dot x  &=&  (2E)^{1/2}
 \frac{ {\rm cn}\left( \phi \;   ;k \right)}
      { {\rm dn}^2\left( \phi  \;  ; k  \right)} ,
\label{dufv}
 \end{eqnarray}
where  the functions sd, cn and dn are Jacobi elliptic functions
\cite{abram}. The elliptic modulus $k$ 
 is a function of  the energy and is  given by
  \begin{equation}
 k^2 = \frac{\sqrt{4E + 1} - 1}{2 \sqrt{4E + 1}} .
 \label{modulus}
\end{equation} 
 In the $(E,\phi)$ coordinate system, the dynamical
 equations for the  Duffing  oscillator,  
 without noise and dissipation, become  simply 
   \begin{eqnarray}
 \dot E &=&   0  \, , \nonumber \\
 \dot\phi &=& (4E + 1)^{\frac{1}{4}}  \, .
\end{eqnarray}

 We reintroduce  the multiplicative  noise and the dissipation terms,
  and  obtain a system of coupled stochastic equations
 that describes the evolution of the energy and angle variables:
  \begin{eqnarray}
 \dot E &=&  J_E(E,\phi) + D_E(E,\phi)  \,\Xi(t) \,\,,
 \label{duffEJD}    \\
 \dot \phi &=&  J_{\phi}(E,\phi) + D_{\phi}(E,\phi) 
  \,\Xi(t) \,\,,
 \label{duffangleJD}
\end{eqnarray}
where the drift and diffusion coefficients $J_E$, $J_\phi$, $D_E$ and 
$D_\phi$ are defined by (see \cite{philkir1} for details of the derivation):
  \begin{eqnarray}
&&J_E(E,\phi) =  - 2 \, \alpha \, E \, \frac{ {\rm cn}^2(\phi;k)}
{ {\rm dn}^4(\phi;k)}\\
&&D_E(E,\phi) =
  \frac { 2E } { (4E + 1)^{\frac{1}{4}}} \,
\frac{{\rm sn}(\phi;k) {\rm cn}(\phi;k)}{{\rm dn}^3(\phi;k)}  \\
&&J_{\phi}(E,\phi) =  (4E + 1)^{\frac{1}{4}}
+ \alpha \,  \frac{2E + 1}{4E + 1} \,
\frac{{\rm sn}(\phi;k) {\rm cn}(\phi;k)}{{\rm dn}^3(\phi;k)} \\
&&D_{\phi}(E,\phi) =
 -  \frac{(2E + 1)}{ (4E + 1)^{\frac{5}{4}}}
\, {\rm sd}^2(\phi;k) \nonumber  \\
&& -   \frac{E}{ (4E + 1)^{\frac{7}{4}}} 
{\rm sd}^4(\phi;k)
 +   \frac{\alpha  \,E }{ (4E +1)^{\frac{3}{2}}} \,
\frac{ {\rm cn}(\phi;k) {\rm sd}^3(\phi;k)}{{\rm dn}^2(\phi;k)}
\nonumber\\
&& +   \frac{E}{ (4E + 1)^{\frac{7}{4}}} \;
  \frac{ {\rm sn}(\phi;k) {\rm cn}(\phi;k)}{{\rm dn}^3(\phi;k)} \,
   \int_0^\phi  {\rm sd}^2(\theta;k) d\theta   \nonumber \\
&& -  \frac{\alpha  \,E }{ (4E +1)^{\frac{3}{2}}} \,
 \frac{ {\rm cn}^2(\phi;k)}{{\rm dn}^4(\phi;k)} \,
   \int_0^\phi  {\rm sd}^2(\theta;k) d\theta   .  \label{duffangle}
\end{eqnarray}

\section{Analysis of the Fokker-Planck equation}
\label{sec:pdf}

  Using  the same notations, 
 the  Fokker-Planck equation that governs the time evolution of
 the probability distribution function $P_t(E,\phi)$ associated
 with the stochastic differential  system
 (\ref{duffEJD})-(\ref{duffangleJD}) reads \cite{vankampen}
\begin{eqnarray}
 \label{FPEphi}   
\partial_t P_t(E,\phi)  &=& - \partial_{E}\left(  J_E P \right)
   -  \partial_{\phi}\left(  J_{\phi} P  \right) \\
  +   \frac{ \Delta}{2} && \left[\right. 
 \partial_{E} \left( D_E\partial_{E} (D_E P ) \right) +
 \partial_{E}  \left( D_E \partial_{\phi}  
(D_{\phi} P )\right) \nonumber \\ 
&& +   \partial_{\phi} 
  \left( D_{\phi}\partial_{E} (D_E P ) \right)
 + \partial_{\phi}
   \left( D_{\phi}\partial_{\phi}(  D_{\phi} P ) \right)
 \left. \right]. \nonumber
\end{eqnarray}
The goal of this section is to determine the conditions
under which there exist stationary 
solutions $P_{\rm{stat}}(E, \phi)$ of Eq.~(\ref{FPEphi})
with a non-zero mean energy. Such an extended PDF
describes an oscillatory asymptotic state.

Obtaining an analytical expression of $P_{\rm{stat}}(E, \phi)$ 
in the general case is a daunting task.
In the following, we shall perform a local analysis of  
Eq.~(\ref{FPEphi}) for small and large values of $E$.
In these limiting cases, the expressions of  the  drift and 
diffusion coefficients  take  simpler  forms. We may write, 
in full generality,
\begin{equation}
   P_{\rm{stat}}(E, \phi) = 
P_{\rm{stat}}(\phi | E) \, {\tilde P}_{\rm{stat}}(E)    \, ,
\label{probcond}
\end{equation}
where $P_{\rm{stat}}(\phi | E)$  represents, in the stationary state,
the conditional probability of the angle $\phi$ at a fixed value of 
the energy $E$. Assuming    the conditional probability  
$P_{\rm{stat}}(\phi | E)$  to be independent of $E$ when $E \ll 1$ and 
$E \gg 1$ (respectively Secs.~\ref{sec:pdf:small} and
\ref{sec:pdf:large}), we shall derive the local behaviour 
of a stationary  solution of the Fokker-Planck equation~(\ref{FPEphi}).
This solution is a legitimate PDF when it is normalizable: the normalizability
condition provides the location of the bifurcation \cite{graham}.

\subsection{Small $E$ limit: the linearized stochastic oscillator}
\label{sec:pdf:small}

  When the oscillator's mechanical energy $E$ is small,
  the position $x(t)$ and the velocity $\dot x(t)$ are also  small:
 nonlinear terms may be neglected. The Duffing oscillator simplifies
 to a  linear (harmonic) oscillator:
 \begin{equation}
   \ddot x + \alpha \dot x + x  = x \; \Xi(t) 
\label{lineaire}
\end{equation}
with $ \langle \Xi(t) \Xi (t') \rangle = \Delta \; \delta(t - t')$.
The energy now  reads  $E = \frac{1}{2} \, {\dot x}^2 + \frac{1}{2} \, x^2$,
and Eqs.~(\ref{dufx}) and (\ref{dufv}) reduce to 
 \begin{eqnarray}
         x &=& \sqrt{2E} \sin \phi  \, , \label{enanglelin1} \\
      \dot x &=& \sqrt{2E} \cos  \phi  \, . \label{enanglelin}
  \end{eqnarray}
In the small $E$ limit, the elliptic modulus $k$ goes to zero,  the
elliptic functions reduce to circular functions \cite{abram}, and 
 the drift and diffusion
coefficients simplify to
\begin{eqnarray}
J_E(E,\phi) &=& -2 \, \alpha \, E \, \cos^2\phi  \,    \label{lim2JE}\\
D_E(E,\phi) &=& 2\, E \, \sin\phi  \cos \phi   \, , \label{lim2DE} \\ 
J_{\phi}(E,\phi) &=& 1  + \alpha  \sin \phi  \cos \phi \, ,\label{lim2Jphi} \\ 
D_{\phi}(E,\phi) &=& - \sin^2 \phi \, . \label{lim2Dphi}
\end{eqnarray}
Substituting the  expressions (\ref{lim2JE})--(\ref{lim2Dphi})
in   Eq.~(\ref{FPEphi}) and  integrating 
over $E$,  we obtain an independent Fokker-Planck equation for 
the marginal PDF ${\tilde P}_{\rm{t}}(\phi)$:
\begin{eqnarray}
  \label{FPlin2}
\partial_t {\tilde P}_t(\phi) &=&  - \partial_{\phi} \left[
  (1 + \alpha \sin\phi \cos\phi)\,   {\tilde P}_t(\phi) \right] \nonumber\\
&&  +     \frac{\Delta}{2} \partial_{\phi} \left[ \sin^2\phi \, 
 \partial_{\phi}  (\sin^2\phi \, {\tilde P}_t(\phi)) \right] \,.
\end{eqnarray}
 An explicit formula for   the  stationary angular
  measure  ${\tilde P}_{\rm{stat}}(\phi)$, solution of
       Eq.~(\ref{FPlin2}), is  derived 
in Appendix~\ref{sec:pdfangle}.

  We now make the  hypothesis
  that in the small $E$ limit,  the stationary  conditional  measure
   $P_{\rm{stat}}(\phi | E )$  becomes independent  of   $E$
 and  is  identical to  the stationary
 angular  distribution    ${\tilde P}_{\rm{stat}}(\phi)$
        of   the harmonic oscillator:  
\begin{equation}
  \hbox{ for }
   \,\,\,\,\,    E \ll  1 \,, \,\,\,\,\,    P_{\rm{stat}}(\phi | E ) = 
  {\tilde P}_{\rm{stat}}(\phi)   \, .    \label{hypfactor}
\end{equation}
Equation~(\ref{probcond}) becomes
 \begin{equation}
   P_{\rm{stat}}(E, \phi) =  {\tilde P}_{\rm{stat}}(E)
   {\tilde P}_{\rm{stat}}(\phi)   \, ,
\label{factor}
\end{equation}
i.e., we assume that the stationary PDF factorizes when $E \ll 1$.
An exactly solvable equation for  
${\tilde P}_{\rm{stat}}(E)$ can now be obtained by
inserting expressions (\ref{lim2JE})--(\ref{lim2Dphi})
and the identity~(\ref{factor}) in the Fokker-Planck  
equation~(\ref{FPEphi}) and then averaging  over $\phi$:
\begin{eqnarray}
& \frac{\Delta}{2} \, \left\langle \sin^2\phi \cos^2\phi \right\rangle \,
  \partial_E \left( E  {\tilde P}_{\rm{stat}}(E) \right) =
\label{FPEmargpetit} \\
&\left( \frac{\Delta}{2}  \left\langle \sin^2\phi \,
 ( \sin^2\phi -  \cos^2\phi) \right\rangle  - \alpha
 \left\langle  \cos^2\phi \right\rangle  \right) {\tilde P}_{\rm{stat}}(E) 
 \nonumber \,.  
\end{eqnarray}
Expectation values denoted by brackets $\langle \, \rangle$ 
are calculated using the stationary angular measure
   ${\tilde P}_{\rm{stat}}(\phi)$.
Introducing the notations:
 \begin{eqnarray}
  \mu(\alpha, \Delta) &=&  \frac{\Delta}{2} \, 
\left\langle \sin^2\phi \cos^2\phi \right\rangle  \, ,  \label{defmu}   \\
  \Lambda(\alpha, \Delta)  &=&  \frac{\Delta}{2}  \left\langle \sin^2\phi \,
 ( \sin^2\phi -  \cos^2\phi) \right\rangle  - \alpha
 \left\langle  \cos^2\phi \right\rangle  \, ,
 \label{deflyap1}
\end{eqnarray} 
Eq.~(\ref{FPEmargpetit}) becomes:
 \begin{equation}
\mu \,  \partial_E \left( E  {\tilde P}_{\rm{stat}}(E) \right) 
=  \Lambda \, {\tilde P}_{\rm{stat}}(E) 
 \,,  \label{FPEmargpetit2}
\end{equation}
and admits the solution:
\begin{equation}
 \label{pdfEpetit}
  {\tilde P}_{\rm{stat}}(E) \propto   E^{ \frac{\Lambda }{\mu }
  - 1 }  \,\,\,\,\,    \hbox{ for }  \,\,\,\, \, E \ll  1 \,.
\end{equation}
Thus, for $E \ll 1$, the PDF ${\tilde P}_{\rm{stat}}(E)$ behaves as a 
power law of the energy.

The coefficient $\mu$ is always positive, but the sign of  $\Lambda$ 
is a function of the parameters $\alpha$ and $\Delta$. 
 For negative values of $\Lambda$,  the stationary solution
 (\ref{pdfEpetit}) is not normalizable and cannot represent a PDF;
 the only stationary  PDF solution  of (\ref{FPEphi})
is then a delta function centered  at  the origin:
\begin{equation}
 \label{pdfEpetitdelta}
  {\tilde P}_{\rm{stat}}(E) =   \delta(E)
   \,\,\,\,\,    \hbox{ for }  \,\,\,\, \, \Lambda < 0 \, .
\end{equation}
 Thus for negative values of  $\Lambda$,  the origin is a global
attractor for the stochastic nonlinear oscillator.
Non-trivial stationary solutions exist only for $\Lambda > 0$:
the bifurcation threshold corresponds to values of the parameters
 for which  $\Lambda(\alpha, \Delta) = 0$.

\subsection{Large  $E$ limit}
\label{sec:pdf:large}

 For very large values of $E$, the elliptic modulus $k$,
  defined in Eq.~(\ref{modulus}),   tends
  to   $1/\sqrt{2}$, and 
  the drift and diffusion coefficients become 
   \begin{eqnarray}
J_E(E,\phi) &=& -2 \, \alpha \, E \, \frac{\rm{cn}^2(\phi)}{\rm{dn}^4(\phi)},
\label{limJE}                  \\ 
D_E(E,\phi) &=& \sqrt{2} \, E^{3/4} \,
\frac{\rm{sn}(\phi) \, \rm{cn}(\phi)}{\rm{dn}^3(\phi)} ,
\label{limDE}                  \\ 
J_{\phi}(E,\phi) &=& (4E)^{1/4} + \frac{\alpha}{2}\,
  \frac{\rm{sn}(\phi) \, \rm{cn}(\phi)}{\rm{dn}^3(\phi)},
\label{limJphi}                  \\ 
D_{\phi}(E,\phi) &=& - \frac{1} {2\sqrt{2} E^{1/4}}  \, \rm{sd}^2(\phi),
\label{limDphi}
 \end{eqnarray}
in agreement with Eqs.~(3) and (4) of \cite{philkir2}
 (we have omitted the elliptic modulus $1/\sqrt{2}$ for
  sake of  simplicity).
  Using  expressions (\ref{limJE}) to (\ref{limDphi})
  for the drift and diffusion
 coefficients, the stochastic equations 
 (\ref{duffEJD}) and (\ref{duffangleJD})  reduce,
  in the large $E$ limit,  to 
\begin{eqnarray}
 \dot{E} & = & -2 \, \alpha \, E \,
  \frac{\rm{cn}^2(\phi)}{\rm{dn}^4(\phi)}
 + \sqrt{2} \, E^{3/4} \,
\frac{\rm{sn}(\phi) \,
  \rm{cn}(\phi)}{\rm{dn}^3(\phi)}\, \Xi(t)   \label{grandE}   \\ 
 \dot{\phi} & =  & (4E)^{1/4} + \frac{\alpha}{2} 
  \frac{\rm{sn}(\phi) \, \rm{cn}(\phi)}{\rm{dn}^3(\phi)}
    - \frac{\sqrt{2}} {4 E^{1/4}}  \rm{sd}^2(\phi)\,  \Xi(t) 
 \label{grandphi}
 \end{eqnarray} 
In terms of the variable  $\Omega = E^{1/4},$ 
  Eq.~(\ref{grandE})  becomes  linear and has a form 
 similar to  Ornstein-Uhlenbeck's  equation.
 Because  the elliptic functions are bounded,  
  $\Omega$    saturates to a value of the order
  of $\frac{\Delta}{\alpha}$. Thus, we deduce  from Eq.~(\ref{grandphi})  
  that  the phase $\phi$
 grows linearly with time. Hence,   $\phi$ is a fast variable
 as compared to $\Omega$.
 Integrating  out this  fast variable  from   the
Fokker-Planck equation leads to 
 an effective stochastic dynamics for $\Omega$, 
  as explained in detail in   \cite{philkir1}.  
An equivalent formulation   is to assume  
  that, in the large $E$ limit, 
 the  stationary conditional probability  $P_{\rm{stat}}(\phi | E)$, 
  defined in Eq.~(\ref{probcond}), becomes   uniform in $\phi$  and 
  independent  of  $E$, i.e.,
  \begin{equation}
 \hbox{ for }  \,\,\,\, \, E \gg 1 \, , \,\,\,\,\,
P_{\rm{stat}}(\phi \,  | \,  E)  \sim 1 \, .
  \end{equation}
 Averaging the dynamics  over $\phi$  
 yields  the following expression 
 for the stationary  PDF  \cite{lindenberg,philkir2}: 
\begin{equation}
  \hbox{ for }   \, E \gg 1 , \,
  {\tilde P}_{\rm{stat}}(E) \propto  E^{-1/4}
\exp \left\{ - \frac{8 \, \alpha \, E^{1/2}}{ 3 \,{\mathcal M} \, {\Delta}  }
 \right\} \, , 
\label{pdfEgrand}
\end{equation}
 with $ {\mathcal M} = \frac{8}{5 \pi^2} \Gamma\left( 3/4 \right)^4 \simeq
 0.3655$,
 $\Gamma$  being  the Eulerian factorial function.
A stationary solution of (\ref{FPEphi}) decays as a stretched exponential
 and is always integrable at infinity
irrespective of the values of $\alpha$ and $\Delta$.
   
\section{Phase Diagram of the Duffing oscillator}
\label{sec:lyapunov}

In the long time limit, the energy of a harmonic oscillator 
with parametric noise (\ref{lineaire})  varies exponentially with time 
\cite{bourret}. This behaviour is characterized by a 
Lyapunov exponent, defined as 
$\lim_{t \to \infty}  \; \frac{1}{2t} \,    \overline{ \log E  }$,
where the overline  means averaging over $P_t(E, \phi)$ \cite{hansel}.
[The factor $1/2$ is included for consistency with the usual definition 
of the Lyapunov exponent in terms of the position $x(t)$.]

In Appendix~\ref{sec:identity}, we prove that the coefficient  
$\Lambda(\alpha,\Delta)$ defined in  Eq.~(\ref{deflyap1})
is equal to the Lyapunov exponent:
\begin{equation}
\Lambda(\alpha,\Delta) =
 \lim_{t \to \infty}  \; \frac{1}{2t} \,    \overline{ \log E  } \, .
 \label{defLyap1}
\end{equation}

\begin{figure}[ht]
\includegraphics*[width=0.75\columnwidth]{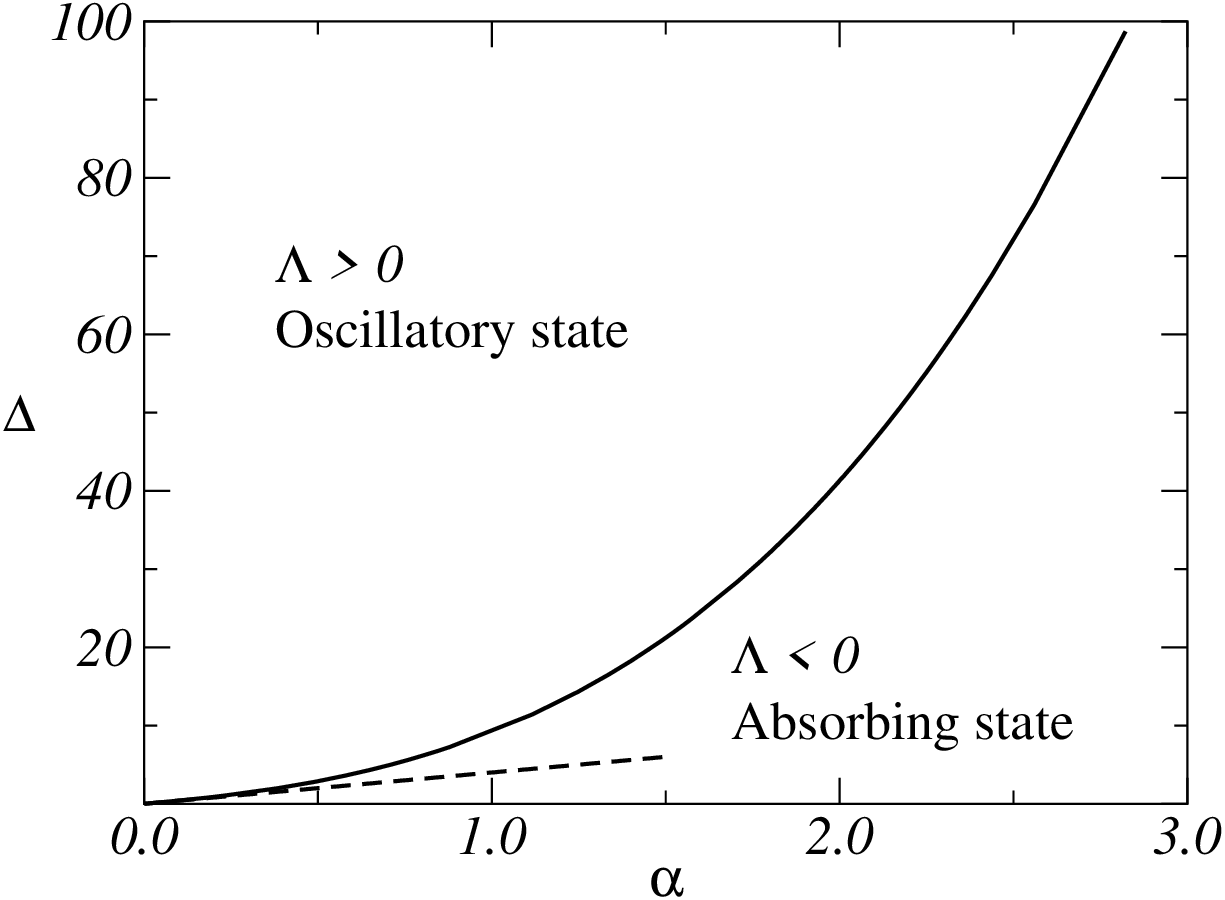}
\caption{\label{fig:lyap:diag}
Phase diagram of the Duffing oscillator with multiplicative noise.
The critical curve separates an absorbing and an oscillatory phase
in the $(\alpha,\Delta)$ plane. The dashed line represents the 
asymptotic approximation,  Eq.~(\ref{smalla}).}
\end{figure}

For the (nonlinear) Duffing oscillator, we found in Section~\ref{sec:pdf}
that a non-trivial stationary 
solution $P_{\rm{stat}}(E, \phi)$ of Eq.~(\ref{FPEphi})
exists only when the Lyapunov exponent  $\Lambda(\alpha, \Delta)$
of the associated linear oscillator is positive.
This extended PDF describes an oscillatory asymptotic 
state for the Duffing oscillator, as opposed to the
absorbing state associated with the delta function (\ref{pdfEpetitdelta}).
The location of the stochastic bifurcation threshold of the Duffing 
oscillator, which separates the two states, is given by the curve  
$\Lambda(\alpha, \Delta) = 0$. The following  explicit formula 
 is derived in Appendix~\ref{sec:pdfangle}:
\begin{equation}
  \Lambda(\alpha, \Delta)  = \frac{1}{2} \left\{  \frac {\int_0^{+\infty} 
\mathrm{d}u \; {\sqrt u} \;
  e^{ -\frac{2}{\Delta} \left(  ( 1 - \frac{\alpha^2}{4} ) u
   + \frac{u^3}{12} \right) }  }
 {\int_0^{+\infty}  \frac{\mathrm{d}u}{\sqrt u}
  e^{-\frac{2}{\Delta} \left(  ( 1 - \frac{\alpha^2}{4} ) u
   + \frac{u^3}{12} \right) } }
   -  \alpha \right\}   \, .       
\label{Lyapunov}
\end{equation}

  The transition line is
  determined  by  the relation between $\alpha$ and $\Delta$ for which
 the Lyapunov exponent vanishes. For a given value
 of $\Delta$, the  critical value of the damping 
$\alpha = \alpha_c(\Delta)$  satisfies 
\begin{equation}
\alpha_c   =  {\displaystyle { \frac {\int_0^{+\infty} 
\mathrm{d}u \; {\sqrt u} \; 
  \exp\left\{-\frac{2}{\Delta} \left(  ( 1 - \frac{\alpha_c^2}{4} ) u
   + \frac{u^3}{12} \right) \right\}  }
 {\int_0^{+\infty} \frac{\mathrm{d}u}{\sqrt u}
  \exp\left\{-\frac{2}{\Delta} \left(  ( 1 - \frac{\alpha_c^2}{4} ) u
   + \frac{u^3}{12} \right) \right\}  } }. }
\label{alphac}
\end{equation} 
The critical curve 
$\alpha = \alpha_c(\Delta)$ [Eq.~(\ref{alphac})] 
is represented in Fig.~\ref{fig:lyap:diag}.
It separates two regions in parameter space: for $ \alpha <  \alpha_c$
 (resp. $\alpha >  \alpha_c$)
 the Lyapunov exponent is positive (resp. negative),
the stationary PDF is an extended function (resp. a delta
distribution) of the energy, the origin is unstable (resp. stable).
For a detailed analytical study, we must now distinguish three cases 
  corresponding
 to an underdamped ($\alpha_c < 2$), critically damped  ($\alpha_c =2$)
and overdamped ($ \alpha_c > 2$) oscillator.

\subsection{Underdamped oscillator}
\label{sec:lin:diag:under}

The case $\alpha_c < 2$  can formally  be 
mapped onto that of a linear oscillator with no damping.
In terms of the variables 
$v = u \left[1 - \frac{\alpha_c^2}{4} \right]^{-1/2}$ and 
$\Delta_< = \Delta \left[1 - \frac{\alpha_c^2}{4}\right]^{-3/2}$,
Eq.~(\ref{alphac}) reads
\begin{eqnarray}
\alpha_c &=&  \left[1 - \frac{\alpha_c^2}{4}\right]^{1/2} \;
    \frac {   \int_0^{+\infty} \mathrm{d}v \, {\sqrt v} \,
  \exp\left\{-\frac{2}{\Delta_<} \left( v+\frac{v^3}{12} \right) \right\} }
 {\int_0^{+\infty} \frac{\mathrm{d}v}{\sqrt v} \exp\left\{-\frac{2}{\Delta_<}  
\left(  v    + \frac{v^3}{12} \right) \right\}  }  \nonumber \\
   &=&  \frac{\Delta_< }{2} \;
\left[1 - \frac{\alpha_c^2}{4}\right]^{1/2}    \;
 \frac { \sum_{n = 0}^{\infty}  \frac{(-1)^n }{n!} \,
\Gamma\left(3n + \frac{3}{2}\right) \,
\left(\frac{\Delta_<^2 }{48}\right)^n}
{  \sum_{n = 0}^{\infty} \frac{(-1)^n}{n!} \,
\Gamma\left(3n + \frac{1}{2}\right) \,
 \left(  \frac{\Delta_<^2 }{48}\right)^n }    \, . \nonumber
\end{eqnarray} 
 Retaining   only  the lowest order term in this series,
  we obtain  the limiting   behaviour  of   the critical
 curve $\alpha_c(\Delta)$,  for small values of $\alpha$ and $\Delta$:
 \begin{equation}
        \alpha_c  = \frac{\Delta  }{4} .
\label{smalla}
\end{equation} 
 The change of variables used here breaks down at $\alpha_c = 2$,
 we shall therefore consider this case separately.

\subsection{Critically damped oscillator}
\label{sec:lin:diag:crit}

  For   a critically damped oscillator,
 we substitute $ \alpha_c =  2 $  in Eq.~(\ref{alphac}) and  obtain
 \begin{equation}
 2 = \frac {\int_0^{+\infty} \mathrm{d}u \, {\sqrt u}\,
  \exp\left(-  \frac{u^3}{6 \Delta_c}  \right)   }
 {\int_0^{+\infty} \frac{\mathrm{d}u}{\sqrt u} \,
 \exp\left(-  \frac{u^3}{6 \Delta_c}  \right)  }
  = ( 6 \Delta_c )^{1/3} \, \frac {\Gamma\left( \frac{1}{2} \right)}
{\Gamma\left(\frac{1}{6} \right)} \, .
\label{dcrit}
\end{equation} 
 Thus,   the critical
noise amplitude $\Delta_c$   for  $ \alpha_c =  2 $ is 
  \begin{equation} 
 \Delta_c = \frac{4}{3} \;  \frac{ \Gamma\left(\frac{1}{6} \right)^3}
{ \pi^{3/2}} \simeq   41.2969..., 
 \label{dcrit2}
\end{equation} 
and the expression~(\ref{Lyapunov}) for the Lyapunov exponent reduces to
\begin{equation}
  \label{eq:Lyap:alpha2}
  \Lambda =  \left(\frac{\Delta}{\Delta_c} \right)^{1/3} - 1 \, .
\end{equation}

\subsection{Overdamped oscillator}
\label{sec:lin:diag:over}

  For  $\alpha_c > 2$,  we define
  $v = u \left[ \frac{\alpha_c^2}{4} - 1 \right]^{-1/2} $  and 
 $\Delta_> = \Delta \left[ \frac{\alpha_c^2}{4} - 1 \right]^{-3/2}$. 
  Equation~(\ref{alphac}) then becomes 
\begin{eqnarray}
\alpha_c \, \left[\frac{\alpha_c^2}{4}- 1 \right]^{-1/2}   
  &=&    \frac {   \int_0^{+\infty} \mathrm{d}v \, {\sqrt v} \,
e^{-\frac{2}{\Delta_>} \left(- v +\frac{v^3}{12} \right) } }
 {\int_0^{+\infty} 
\frac{\mathrm{d}v}{\sqrt v} e^{-\frac{2}{\Delta_>}  
\left(  -v    + \frac{v^3}{12} \right) }  }  \label{alphacseries2} \\
   &=& (6{\Delta_> })^{1/3} \;
 \frac  { \sum_{n = 0}^{\infty} 
\frac{1}{n!} \Gamma\left(\frac{2n+3}{6}\right) 
\left(\frac{48}{\Delta_>^2 }\right)^{n/3 } }
{  \sum_{n = 0}^{\infty}  \frac{1}{n!} \,
\Gamma\left(\frac{2n+1}{6}\right) 
\left(\frac{48}{\Delta_>^2 }\right)^{n/3 } }  \nonumber  
\end{eqnarray}  

 When $\alpha_c \to \infty$, the left hand side  of  Eq.~(\ref{alphacseries2})
the previous equation  converges to 2  and 
   ${\Delta_>}$  tends to  a limiting value ${\cal K}$  that  
we computed numerically:  ${\cal K}  \simeq 28.337$.
 Thus, in the large  $\alpha_c$ limit, we   find  
 the following relation   between $\alpha_c$ and $\Delta$
\begin{equation}
 \Delta  \simeq  {\cal K} \,\left[\frac{\alpha_c^2}{4}- 1 \right]^{3/2} 
 \simeq \frac{  {\cal K} }{8} \,  \alpha_c^3 \simeq 3.54 \, \alpha_c^3 .
\label{largea}
  \end{equation} 

When $\alpha_c  \to  2^{+}$, the left hand side of Eq.~(\ref{alphacseries2})
increases to  infinity.  Retaining only   the leading order term 
  of the series, we obtain 
\begin{equation}
2 \, \left[\frac{\alpha_c^2}{4}- 1 \right]^{-1/2}=
    (6{\Delta_> })^{1/3}  \frac{\Gamma(1/2)}{\Gamma(1/6)} \, .
\end{equation} 
 This equation  is consistent with Eq.~(\ref{dcrit2}).

\section{Physical observables}
\label{sec:scaling}

The goal of this section is  to determine how  physical observables,
such as the energy and the position of the oscillator,  scale
near  the  bifurcation threshold as well as far from it. 
The agreement between analytical calculations
and numerical simulations will validate the assumptions 
made in Section~\ref{sec:pdf} concerning the behaviour of the
conditional PDF $P_{\rm{stat}}(\phi | E)$ for small and large values of $E$.

Let us first of all review the results obtained in Section~\ref{sec:pdf}.
For parameter values such that $\Lambda < 0$, we demonstrated that
${\tilde P}_{\rm{stat}}(E)  = \delta(E)$: the origin is a 
global attractor of the dynamics. Conversely, when  $\Lambda > 0$, 
the origin becomes unstable. The stationary measure of the energy
is then an extended function: it behaves as a power law 
for small values  of $E$  [see Eq.~(\ref{pdfEpetit})],
and decreases as a stretched exponential for large values of $E$
[see Eq.~(\ref{pdfEgrand})]:
   \begin{eqnarray}
    \hbox{ for }  \, \, E \ll  1 \, , \,\,\,
  {\tilde P}^{<}_{\rm{stat}}(E) &=&  A  
  E^{ \frac{ \Lambda }{ \mu }
  - 1 }      \,,    \nonumber \\
   \hbox{ for }  \, \, E \gg  1 \, , \,\,\,
  {\tilde P}^{>}_{\rm{stat}}(E) &=& B  E^{-\frac{1}{4}}
\exp \left\{ - \frac{8 \, \alpha \, E^{1/2}}{ 3 \,{\mathcal M} \, {\Delta}  }
 \right\} \, .
 \label{matching}
  \end{eqnarray}
  The precise values of $A$  and $B$ can only
 be obtained by solving  the full Fokker-Planck
  equation~(\ref{FPEphi}), and matching 
  the two asymptotic expressions in the 
  intermediate regime $E \simeq 1$.

\subsection{Behaviour   in the vicinity of the bifurcation}
\label{sec:NL:bif}

\begin{figure}[th]
\includegraphics*[width=0.75\columnwidth]{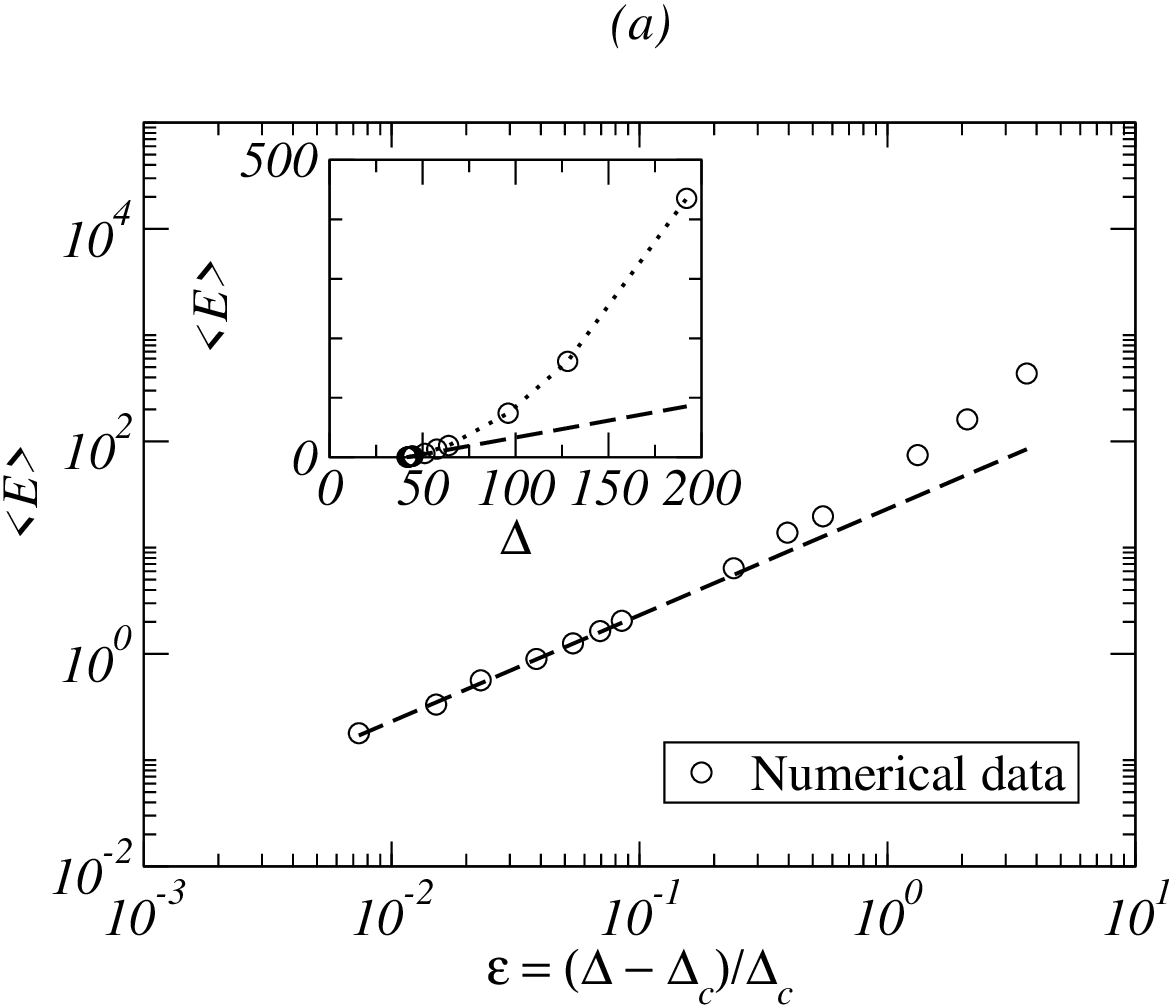}
\bigskip
\includegraphics*[width=0.75\columnwidth]{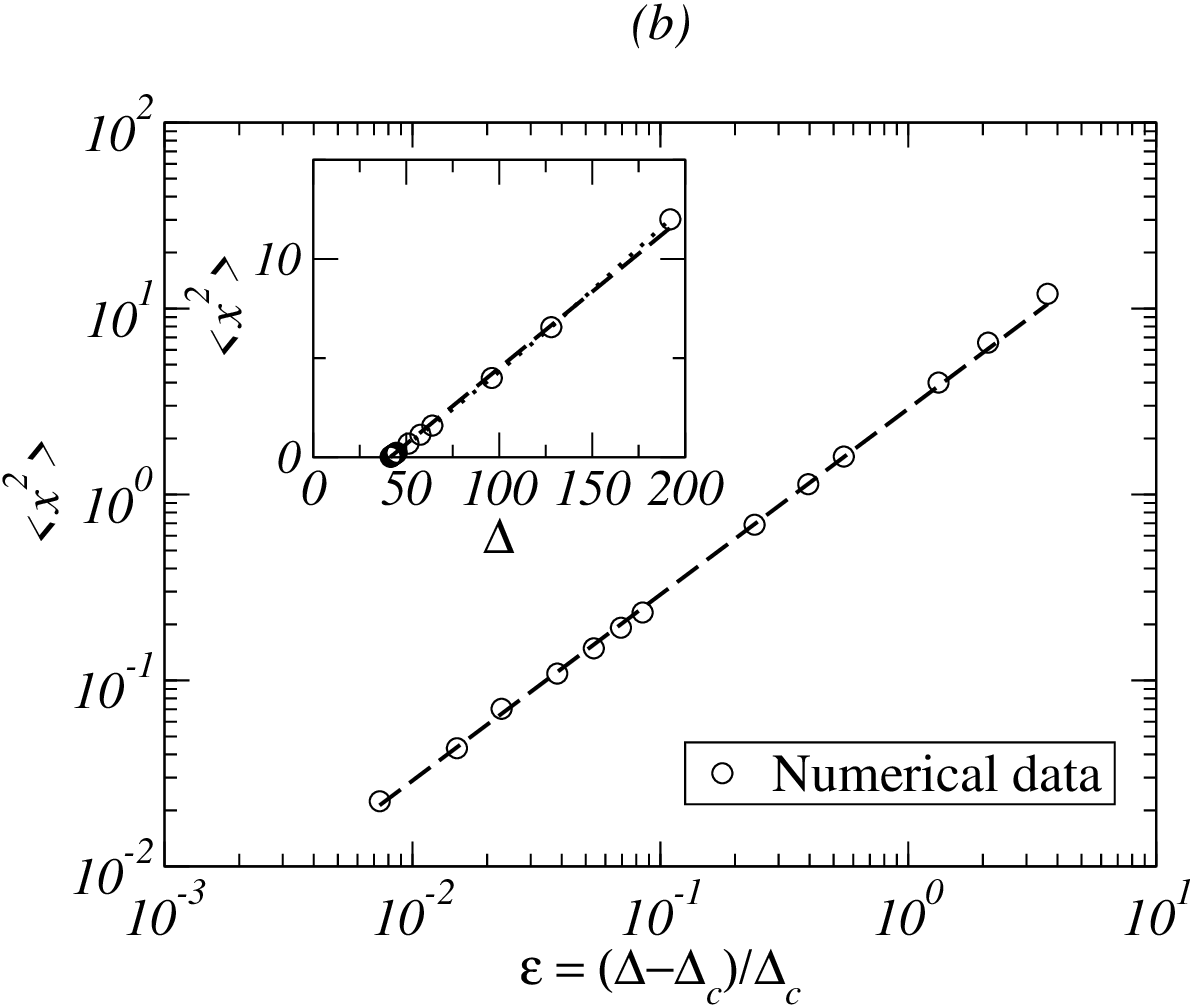}

    \caption{\label{fig:bifn3} Scaling behaviour of physical
observables of the stochastic  Duffing  oscillator
in the vicinity of the bifurcation threshold as a function of
the reduced control parameter 
$\varepsilon = (\Delta - \Delta_c)/\Delta_c$. (a) Average energy; (b) 
Variance of the position. Dashed lines corresponding to a linear function
of $\varepsilon$ are drawn to guide the eye. The corresponding
linear-linear graphs are shown as insets. Eq.~(\ref{dissipDuff})
is integrated numerically for $\alpha = 2$, with a time step
$\delta t = 5 \ 10^{-4}$. The threshold value is given by Eq.~(\ref{dcrit2}):
$\Delta_c(2) \simeq 41.3$. Ensemble averages are computed over
$10^3$ realisations.}
\end{figure}

\begin{figure}[th]
\includegraphics*[width=0.75\columnwidth]{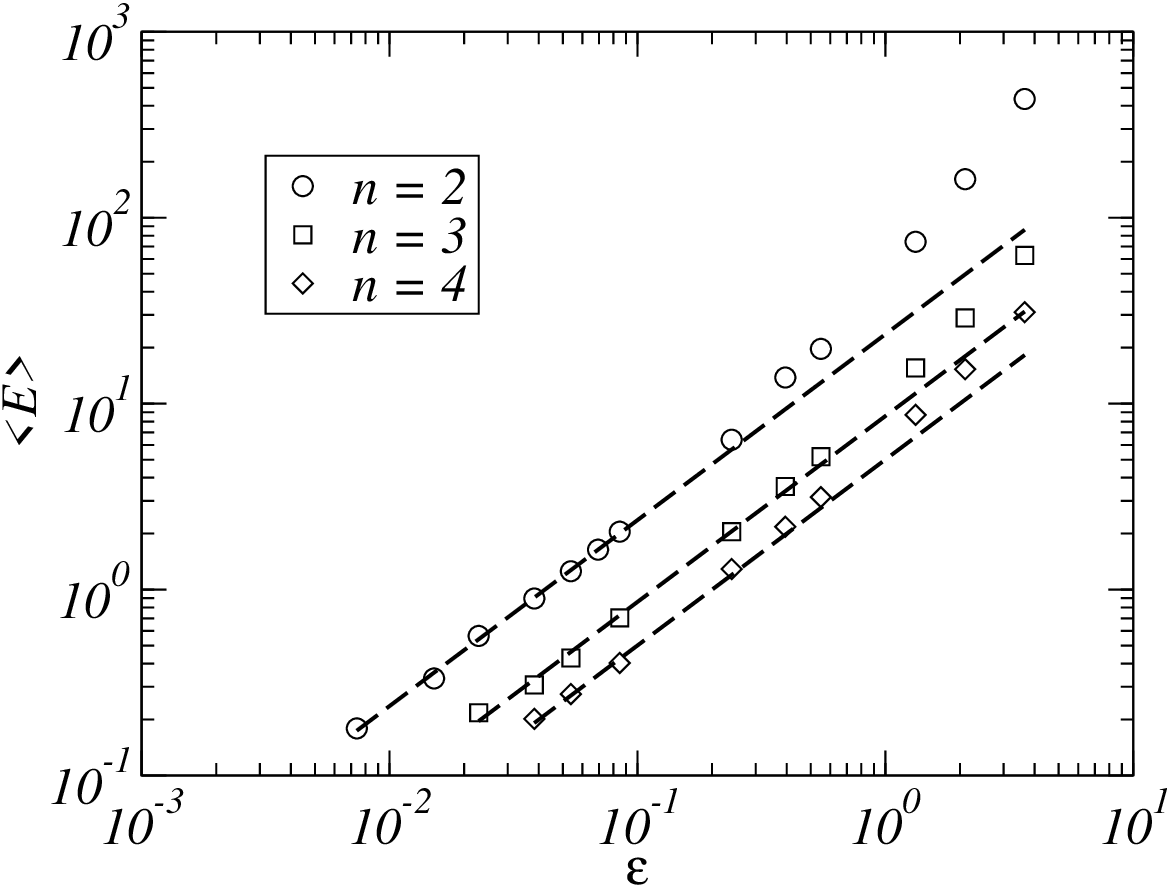}
\caption{\label{fig:bifn} Scaling behaviour of the  mean   energy 
of  the general  nonlinear oscillator in the vicinity of the 
bifurcation threshold as a function of the reduced control 
parameter  $\varepsilon = (\Delta -\Delta_c)/\Delta_c$. Dashed lines
indicating a linear dependence in $\varepsilon$ are drawn to guide the eye.
Eq.~(\ref{eqsto}) is integrated numerically for $\alpha = 2$, 
and with ${\mathcal U}(x) = x^{2n}/(2n)$,   $n = 2$, $3$ and $4$. 
Data obtained for $n = 2$ are the same as in Fig.~\ref{fig:bifn3}. 
The time step is $\delta t = 5 \ 10^{-4}$; ensemble averages are computed over
$10^3$ realisations. }
\end{figure}

   For  a given value of the damping rate $\alpha$,
 let us choose the noise strength  $\Delta$  such
 that  $0 < \Delta -\Delta_c(\alpha) \ll 1$. In this case, 
   the parameters of the  system 
 are tuned just   above the bifurcation threshold  and 
    the Lyapunov exponent  is slightly positive:  
 $0 < \Lambda  \ll 1$.
  The stationary PDF is an extended function but,
  as can be inferred  from   Eq.~(\ref{matching}), 
 most of its  mass  is concentrated in  the vicinity 
 of $E=0$.  Therefore, a good approximation of 
the stationary PDF  is obtained by taking it equal to 
    ${\tilde P}^{<}_{\rm{stat}}(E)$  over  a finite
 interval in the vicinity of 0  (e.g.,   $0 \le E \le E_0$),  
 and  equal to zero     outside  this  interval.  
The mean value of the energy is given by:
\begin{equation}  
  \langle E \rangle_{\mathrm{stat}} \simeq 
 \frac{  \int_0^{E_0} E^{\frac{\Lambda }{\mu}} \mathrm{d} E   } 
      { \int_0^{E_0}  E^{ \frac{\Lambda }{\mu }  - 1 }  \mathrm{d} E  } 
 \propto  E_0 \, \frac { \Lambda }{ \mu} \, .   
  \label{multif1} 
\end{equation}
 If we now take into account that: $(i)$  
$ \Lambda(\alpha, \Delta_c(\alpha)) =0$; and
 $(ii)$  the Lyapunov  exponent and the coefficient $\mu$  are regular
 functions  of the noise  strength $\Delta$ when $\Delta \neq 0$, 
 we obtain from Eq.~(\ref{multif1})
\begin{equation}  
 \langle E \rangle_{\mathrm{stat}} 
\propto \Lambda(\alpha, \Delta) - \Lambda(\alpha, \Delta_c)
\propto (\Delta -\Delta_c)   \, , 
 \label{multif2}
\end{equation} 
where the last identity is obtained by a Taylor expansion 
of $\Lambda$ for $\Delta$ close to $\Delta_c$.
This result is confirmed by numerical simulations 
[See Fig.~\ref{fig:bifn3}.a]. A similar reasoning 
shows that $ \langle E \rangle_{\mathrm{stat}} $ scales
linearly with $\alpha - \alpha_c(\Delta)$ at fixed noise strength $\Delta$.

Let us now consider  the variance of the oscillator's
position close to the bifurcation.
Using Eqs.~(\ref{enanglelin1}) and (\ref{factor}), we find
\begin{eqnarray}
\langle x^2 \rangle_{\mathrm{stat}}  
&=& \iint P_{\mathrm{stat}}(E,\phi) \, 2 E \sin^2 \phi
\; \mathrm{d}E \, \mathrm{d}\phi \nonumber \\
&=& 2 \, \langle \sin^2 \phi \rangle \, 
\langle E \rangle_{\mathrm{stat}} \nonumber \\
&\propto& (\Delta -\Delta_c) \, .    \label{eq:scalingx2}
\end{eqnarray}
The variance of the position scales linearly with the distance to
threshold [See Fig.~\ref{fig:bifn3}.b]. 
Using a similar argument, one also finds that 
$\langle v^2 \rangle_{\mathrm{stat}} \propto (\Delta -\Delta_c)$.

As seen in  Fig.~\ref{fig:bifn}
 the linear scaling (\ref{multif2}) of the energy with respect
  to  the distance to threshold is observed
for  ${\mathcal U}(x) \propto x^{2n}  $, $n = 2$, $3$ and $4$,
where $\Delta_c(\alpha)$ is \emph{always} given by 
 $\Lambda(\alpha, \Delta_c(\alpha)) = 0$.
The threshold is independent of   the nonlinearity 
 involved in the confining  potential ${\mathcal U}$ appearing 
 in Eq.~(\ref{eqsto}), because the  local expression 
 ${\tilde P}^{<}_{\rm{stat}}(E)$  of the stationary PDF
 in the vicinity of $E =  0$ is obtained from the \emph{linearized}
 stochastic differential equation.   The observed dependence of the
constant of proportionality in Eq.~(\ref{multif2}) upon 
the stiffness of the anharmonic potential remains to be understood.

\begin{figure}[ht]
\includegraphics*[width=0.75\columnwidth]{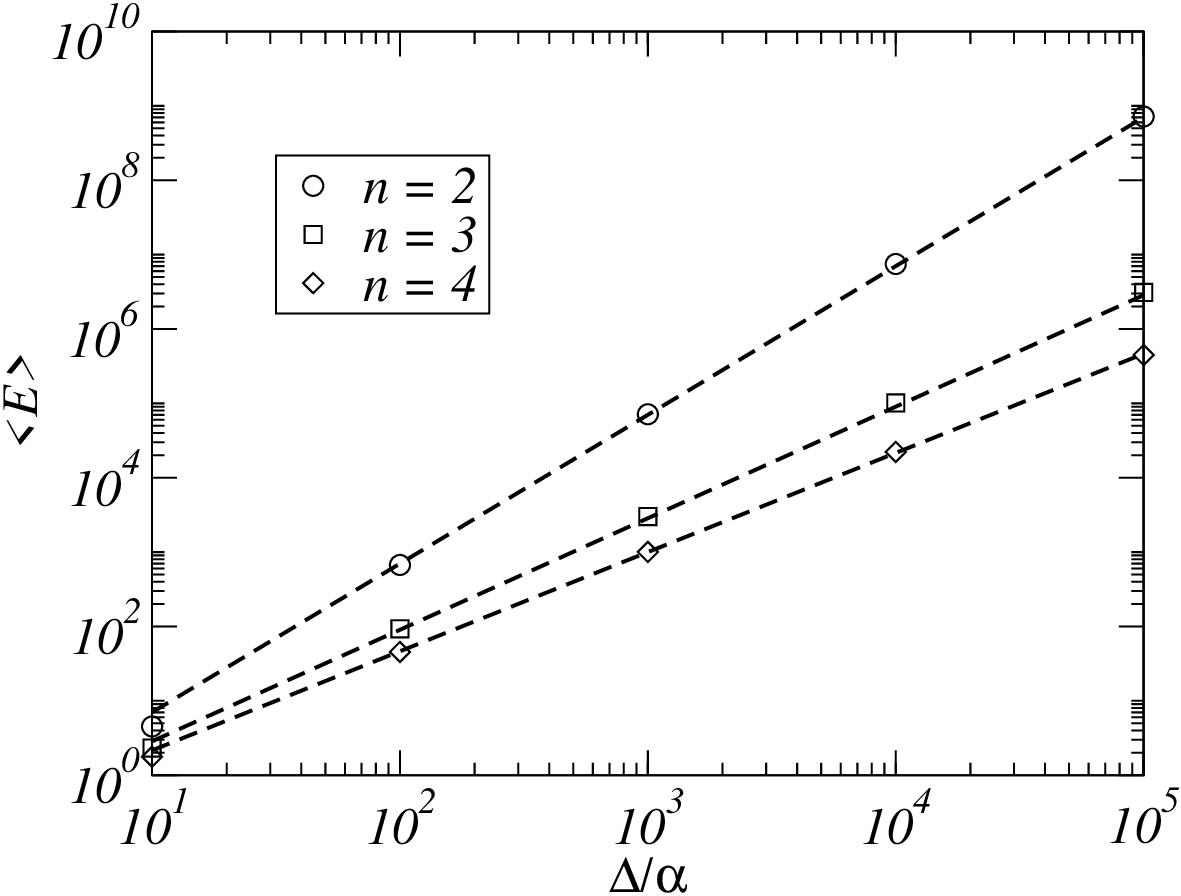}
\caption{\label{fig:loin} Scaling behaviour of the mean energy 
of  the general  nonlinear oscillator far from the bifurcation 
as a function of the parameter  $\Delta/\alpha$. 
Eq.~(\ref{eqsto}) is integrated numerically with 
${\mathcal U}(x) = x^{2n}/(2n)$,   $n = 2$, $3$ and $4$. 
Dashed lines correspond to the predicted scaling (\ref{avgEtn}),
with a scaling exponent $n/(n-1) = 2$, $3/2$ and $4/3$.
The time step is $\delta t = 5 \ 10^{-4}$; ensemble averages are computed over
$10^3$ realisations. }
\end{figure}

 \subsection{Scaling  far from the bifurcation}

  We shall now study the scaling of the  physical observables far from the
 bifurcation point. This situation is encountered,
 for instance,
 when   the damping rate is vanishingly
 small for a given noise strength.  The Lyapunov exponent 
  is then a  finite, strictly positive, number. 
 When $\alpha \to 0$, 
 most of the mass of the stationary PDF is concentrated at large
  values of  $E$  as can be seen from Eq.~(\ref{matching}).
 Thus,  a satisfactory   approximation of 
the stationary PDF  is to assume it to be  identical to the  expression
    ${\tilde P}^{>}_{\rm{stat}}(E)$ for all $E$. 
 After a suitable normalisation, we find 
 \begin{equation}
   {\tilde P}_{\rm{stat}}(E) = \frac{1}{ 2 E \Gamma \left(3/2\right)}
\left( \frac{8 \, \alpha  \, E^{1/2}}{ 3 \,{\mathcal M} \, {\Delta}   }
\right)^{3/2} 
\exp \left\{ - \frac{8 \, \alpha \, E^{1/2}}{ 3 \,{\mathcal M} \, {\Delta} }
 \right\} ,
\label{pdfstat}
\end{equation} 
 with  the numerical factor ${\mathcal M}$ 
   defined in Eq.~(\ref{pdfEgrand}).
 In this low damping regime,  we  recover,
 as expected, the  results derived in Refs.~\cite{philkir1,philkir2}.
 The average energy is readily obtained from Eq.~(\ref{pdfstat})
 \begin{equation}
\langle E \rangle_{\rm{stat}}  = \frac{135 }{256} \, {\mathcal M}^2 \,
\left( \frac{ {\Delta}   }{\alpha }\right)^2 
 \simeq 0.070  \left( \frac{ {\Delta}   }{\alpha }\right)^2  \, .
\label{avgEt}
\end{equation} 
 This result is  
at variance with the usual equilibrium fluc\-tua\-tion-dissipation theorem 
($\langle E \rangle_{\mathrm{eq}} \propto {\Delta}/\alpha $);
 yet this is not  a contradiction since
 Eq.~(\ref{dissipDuff}) models a system  out of  equilibrium.

Far from the bifurcation threshold, the scaling  behaviour 
 depends strongly  on   the stiffness of the potential ${\mathcal U}$. 
If we take ${\mathcal U}(x) \propto x^{2n}  $, Eq.~(\ref{avgEt})
 can be generalized to  \cite{philkir2} 
 \begin{eqnarray}
\langle E \rangle_{\rm{stat}}  = &&
\frac{1}{2n} \;     \frac{\Gamma \left(\frac{3n+1}{2(n-1)} \right)} 
  {\Gamma \left(\frac{n+1}{2(n-1)} \right)}  \label{avgEtn} \\
&&  \left( ( n-1) \; 
     \frac{\Gamma \left(\frac{3}{2n} \right)
 \; \Gamma \left(\frac{3n+1}{2n} \right)} 
  {\Gamma \left(\frac{1}{2n} \right) \; 
 \Gamma \left(\frac{3n+3}{2n} \right)} 
\right)^{\frac{n}{n-1}} \;
 \left( \frac{ {\Delta}   }{\alpha }\right)^{\frac{n}{n-1}} \, .
 \nonumber
 \end{eqnarray}
This prediction is again confirmed by numerical simulations
[See Fig.~\ref{fig:loin}].

 \section{Conclusion}
\label{sec:disc}

    For a  deterministic Duffing  oscillator  with  linear damping,
 the fixed point at the origin $(x = \dot{x} =0)$ is a 
 global attractor: irrespective of the choice of initial conditions, 
 the mechanical energy is totally  dissipated in the long time limit 
and  the system  converges  to the origin.
This picture is no longer valid once the oscillator is 
subject to parametric noise. Indeed, the dissipative  Duffing  
oscillator with parametric white  noise undergoes a noise-induced 
bifurcation between: $(i)$ an absorbing state reminiscent of the 
deterministic system;  and $(ii)$ an oscillatory state, resulting
from a dynamical balance between energy dissipation and 
(noise-induced) energy injection. Noisy oscillations appear
when the fixed point at the origin becomes unstable: this transition 
is thus referred to as a (forward) stochastic Hopf  bifurcation.
Both states may be  characterized by the stationary probability
 distribution function of the oscillator's mechanical energy $E$.
 The PDF of the absorbing state is a  delta-function, whereas
 the oscillatory  state corresponds to an extended PDF.

Our main result concerns the location of the transition point:
a  noise-induced  bifurcation for the Duffing oscillator occurs precisely 
where the Lyapunov exponent of the linear  oscillator with 
parametric noise changes sign. We emphasize that this result
is genuinely non-perturbative. It results from a simple hypothesis 
on the behaviour of the  conditional probability of the angular variable 
at fixed energy, valid for very small values of the energy 
  and for \emph{arbitrary} noise amplitude.
This assumption is physically plausible (i.e. mean-field-like) and
justified by the consequences deduced from it.
 Having identified the Lyapunov exponent  of the stochastic
\emph{linear} oscillator as the important physical quantity, we  
give a simple analytical expression of this exponent as a 
function of the system's parameters, the damping rate and the noise strength.
The location of the transition line that separates the absorbing and
oscillatory states of the (nonlinear) Duffing  oscillator follows
immediately. 

Although we have considered here only a cubic
 oscillator with parametric Gaussian 
  white noise, we believe that the relation
between the sign of the Lyapunov exponent and the
 location of the noise-induced transition holds 
 irrespective  of the nonlinearity and of the nature
 of the noise.   Finally, we demonstrate that
all physical observables scale linearly with respect to the distance
to threshold. Indeed, the argument we use to evaluate the mean energy
$\langle E \rangle$ may be easily generalized to show that 
all moments $\langle E^k \rangle$  also scale linearly,
 suggesting that the concept of critical exponents
 is not relevant here.
One would certainly like to test experimentally whether 
noise-induced transitions may indeed be characterized
by a multifractal order parameter, at variance with the
theory of equilibrium continuous transitions.

\appendix

\section{Some useful identities for the  Lyapunov exponent}
\label{sec:identity}

In this appendix, we  prove that the Lyapunov exponent, defined as 
$\lim_{t \to \infty}  \; \frac{1}{2t} \,    \overline{ \log E  }$,
satisfies the relation~(\ref{defLyap1}). 
  Using  the  `radial' variable  $\rho$  defined by 
\begin{equation}
  \rho = \frac{1}{2} \log \left( 2 E \right)  \,  , \label{defU}
\end{equation}
 we rewrite   Eq.~(\ref{lineaire}) as a
  system of nonlinear  coupled stochastic equations
  \begin{eqnarray}
  \dot \rho &=&  -  \alpha   \cos^2 \phi +  \sin\phi \cos\phi  \, \Xi(t) 
    \, , \label{eqUlin}   \\
   \dot \phi &=&  1 + \alpha \sin\phi \cos\phi - \sin^2\phi\, \Xi(t)
    \, .\label{eqthetalin}
 \end{eqnarray}
  The  Fokker-Planck
  evolution  equation for the joint probability distribution
 $P_t(\rho,\phi)$  is then given by  \cite{vankampen}
 \begin{eqnarray}
 \partial_t P_t& = & \label{FPlin3}
  \alpha   \cos^2 \phi \, \partial_{\rho}  P_t   -  \partial_{\phi} 
\left[ (1 + \alpha \sin\phi \cos\phi)\, P_t \right] \label{FPlin1}  \\ 
&& + \frac{\Delta}{2}  \left\{\right. 
 \sin^2 \phi \cos^2\phi \,  \partial^2_{\rho\rho} P_t
 - \sin\phi \cos\phi \, 
\partial^2_{\rho \phi} \left[\sin^2\phi \, P_t \right]  \nonumber \\
&&  - \partial_{\phi} \left[ \sin^3\phi \cos\phi \,  
\partial_\rho P_t \right]   +   \partial_{\phi} \left[ \sin^2\phi \, 
 \partial_{\phi}  (\sin^2\phi \, P_t \,) \right] \left.\right\} \nonumber
\end{eqnarray}
  From  Eq.~(\ref{FPlin1}) we derive the time evolution
 of the mean value  $\overline{ \rho }$, where  the overline
  indicates  averaging with respect to  
$P_t(\rho,\phi)$. After  suitable integration by parts, we obtain
\begin{equation}
 \frac{ \rm{d}}{ \rm{d} t}\overline{ \rho }  =
      -\alpha \;  \overline{  \cos^2\phi} \,  
 +  \frac{\Delta}{2}  \overline{   \sin^2\phi\,  ( \sin^2\phi -  \cos^2\phi )}
    \, .  \label{evolmoyU}
\end{equation} 
 Similarly,  we deduce the time evolution
 of  $ \overline{ \log | \sin\phi |  }$
 \begin{eqnarray}
  \frac{ \rm{d}}{\rm{d} t}  \overline{ \log | \sin\phi |  }
& = & \overline{ \mathrm{cotan}\, \phi } 
+ \alpha \; \overline{ \cos^2\phi }  \nonumber\\
&& - \frac{\Delta}{2} \; \overline{  \sin^2\phi\,  
( \sin^2\phi -  \cos^2\phi )}
    \, .  \label{evolmoylog}
 \end{eqnarray}
 From Eqs.~(\ref{evolmoyU}) and (\ref{evolmoylog}), we obtain 
 \begin{equation} 
 \frac{ \rm{d}} { \rm{d} t}\overline{ \rho }   =
 \overline{ \mathrm{cotan}\, \phi }    -  
  \frac{ \rm{d}}{\rm{d} t}\overline{ \log | \sin\phi |  } 
  \, .  \label{lienmoys} 
 \end{equation} 
 At large times, the PDF ${  P}_t(E, \phi)$ reaches
 a stationary limit  ${ P}_{\rm{stat}}(E,\phi)$.
 Thus,  when $ t \to \infty $,
  the averages  $\overline{ \log | \sin\phi |  }$   and 
 $\overline{ \mathrm{cotan}\, \phi }$   have  finite limits,
 given by $\langle  \log | \sin\phi | \rangle$  and
 $ \left\langle \mathrm{cotan}\, \phi  \right\rangle $, respectively,
 where the expectation values  denoted by  brackets $\langle \rangle$ 
 are calculated by using the marginal 
 PDF,  ${\tilde P}_{\rm{stat}}(\phi)$.
 Therefore 
 \begin{equation}   
 \lim_{t \to \infty } \frac{ \rm{d}} { \rm{d} t}  \overline{ \rho}    =  
  \left\langle \mathrm{cotan}\, \phi  \right\rangle    \,.
\label{meanrho}
\end{equation}
  From Eqs.~(\ref{defU})~and~(\ref{meanrho}), 
 we derive
\begin{equation}  
 \lim_{t \to \infty}  \frac{1}{2 t} \overline{ \log E }
= \lim_{t \to \infty}  \frac{1}{2} 
\frac{\mathrm{d}}{\mathrm{d} t} \overline{ \log E }
= \lim_{t \to \infty}  \frac{\mathrm{d}}{\mathrm{d} t} 
\overline{ \rho} 
 =    \left\langle \mathrm{cotan}\, \phi  \right\rangle   \, .
\label{formLyap}
\end{equation}
Similarly,  we take the stationary limit of Eq.~(\ref{evolmoylog}) 
and obtain
 \begin{equation}
   \left\langle \mathrm{cotan}\, \phi  \right\rangle  =
- \alpha \; \left\langle \cos^2\phi  \right\rangle 
 +  \frac{\Delta}{2} \;\left\langle  \sin^2\phi\,  
( \sin^2\phi -  \cos^2\phi ) \right\rangle    .
\label{Lyapfin}
 \end{equation}
The right hand side of this equation is nothing but $\Lambda$ [See
Eq.~(\ref{deflyap1})]. This 
concludes the derivation of   Eq.~(\ref{defLyap1}) :
\begin{equation}  
 \lim_{t \to \infty}  \frac{1}{2 t} \overline{ \log E }
 =    \left\langle \mathrm{cotan}\, \phi  \right\rangle  
 =   \Lambda(\alpha, \Delta) \, .
\label{idLyap}
\end{equation}

   \section{Calculation  of the  Lyapunov exponent}
\label{sec:pdfangle}

 Expressions for the  Lyapunov exponent  have been 
  available   for  a long time   
 in the  literature \cite{hansel},
 but they  are rather unwieldy.
 In  a recent  article 
  about the quantum localisation problem \cite{tessieri},
  an  expression for  the 
 localisation length  is derived: it is shown to be identical
 to  the inverse  Lyapunov exponent of the linear  oscillator
  with parametric noise and zero damping. 
 In this appendix, we generalize the calculation
 of \cite{tessieri} to the dissipative case 
 and derive a particularly  simple formula for the Lyapunov exponent.

  Since  $\Lambda = \langle \mathrm{cotan}\phi \rangle  $ 
 [See Eq.~(\ref{idLyap})],  
  we introduce  the auxiliary variable $z$ defined as
\begin{equation}
 z = \dot x/x = \mathrm{cotan}\, \phi  \, .
\label{defz}
\end{equation} 
  Then, equation~(\ref{lineaire})  becomes  
\begin{equation}
\dot z + 1 + \alpha \; z  + z^2 = \Xi(t) \,.
\label{eqz}
 \end{equation} 
 The  stationary Fokker-Planck equation  associated  with Eq.~(\ref{eqz})
  is given by
 \begin{equation}
    \frac{\Delta}{2} \;  \frac{ \rm{d}} { \rm{d} z} P_{\rm{stat}}
 = -( 1 +  \alpha z  +  z^2) P_{\rm{stat}}  + J \,,
\label{StatFPz}
\end{equation} 
where we have introduced the stationary current $J$.
 The solution of this equation   is found  by the method
of the variation of constants:
\begin{equation}
 P_{\rm{stat}}(z) =  \frac{1}{N} \int_{-\infty}^z  \exp\left\{\frac{2}{\Delta} 
  (\Phi(y) - \Phi(z) ) \right\} \mathrm{d}y,
 \label{Pstatz}
\end{equation} 
where $\Phi(y) = y +\frac{\alpha}{2} y^2 + \frac{1}{3} y^3$.
Writing  $\int_{-\infty}^{\infty} P_{\rm{stat}}(z) \; \mathrm{d}z = 1$, we
obtain the normalisation constant $N$:
\begin{equation}
  N = \int_{-\infty}^{+\infty} \mathrm{d}z \int_{-\infty}^z  \mathrm{d}y
\exp\left\{\frac{2}{\Delta}   (\Phi(y) - \Phi(z) ) \right\} .
\label{formN1}
\end{equation} 
  Following \cite{tessieri}, we 
 change the  $y$  variable to $u = z -y$ in Eq.~(\ref{formN1}):
\begin{equation}
   N  =  \int_{-\infty}^{+\infty} \mathrm{d}z \int_0^{+\infty}\mathrm{d}u  \, 
e^{-\frac{2}{\Delta} \big(u  - \frac{\alpha}{2}u^2  + \frac{u^3}{3}
 + z^2 u + z( \alpha u -  u^2) \big) }  \, .
\end{equation} 
  The integral over  $z$ is now Gaussian, and can be evaluated 
 before the integral  over  $u$. Hence, we obtain 
\begin{equation}
   N  = \sqrt{\frac{\pi \Delta}{2}} \; \int_0^{+\infty} 
\frac{\mathrm{d}u}{\sqrt u}  \,
  \exp \left\{ -\frac{2}{\Delta} \,  \left(  ( 1 - \frac{\alpha^2}{4} ) u
   + \frac{u^3}{12} \right) \right\} \, .
\label{Norm}
\end{equation} 

 The use of the identity
 \begin{equation}
  {\tilde P}_{\rm{stat}}(\phi)\rm{d}\phi  =  P_{\rm{stat}}(z) \rm{d}z \, ,
\end{equation} 
  yields the stationary angular measure
 \begin{equation}
  \label{probastatphi}
{\tilde P}_{\rm{stat}}(\phi) = \frac{1}{N \sin^2\phi} 
 \int_{-\infty}^{\rm{cotan}\;\phi}   
\mathrm{d}y   \;
e^{ \frac{2}{\Delta} \left(\Phi(y) - \Phi(\mathrm{cotan}\;\phi) \right)}
 \, .
    \end{equation}
  We deduce that the 
 Lyapunov exponent satifies
 \begin{eqnarray}
\Lambda &=&   
 \int_0^\pi {\tilde P}_{\rm{stat}}(\phi) \, \mathrm{cotan} \phi \,
 \mathrm{d}\phi   \label{eqlambdaz}  \\
&=&  \frac{1}{N}\int_{-\infty}^{+\infty} \mathrm{d}z \; z 
 \int_{-\infty}^z  \exp\left\{\frac{2}{\Delta} 
  (\Phi(y) - \Phi(z) ) \right\} \mathrm{d}y    \, .\nonumber
 \end{eqnarray}
  We again  change the variables  from $(y,z)$ to $(u,z)$
with $u = z -y$, and obtain 
 \begin{equation}  
\Lambda =   \frac{1}{N} 
\int_{-\infty}^{+\infty} \mathrm{d}z \, z \int_0^{+\infty}\mathrm{d}u \,
e^{-\frac{2}{\Delta} \left(u  - \frac{\alpha}{2}u^2  + \frac{u^3}{3}
 + z( \alpha u -  u^2)  + z^2 u   \right) }  
\end{equation} 
  Evaluating the Gaussian integral in $z$  leads to
 the     expression   of  the   Lyapunov exponent given in the text
 \begin{equation}
  \Lambda(\alpha, \Delta)  = \frac{1}{2} \left\{  \frac {\int_0^{+\infty} 
\mathrm{d}u \; {\sqrt u} \;
  e^{ -\frac{2}{\Delta} \left(  ( 1 - \frac{\alpha^2}{4} ) u
   + \frac{u^3}{12} \right) }  }
 {\int_0^{+\infty}  \frac{\mathrm{d}u}{\sqrt u}
  e^{-\frac{2}{\Delta} \left(  ( 1 - \frac{\alpha^2}{4} ) u
   + \frac{u^3}{12} \right) } }
   -  \alpha \right\}   .      
\label{Lyapunovapp}
\end{equation} 
  This formula also appears in 
  recent   mathematical   literature  \cite{imkeller}. 
  It constitutes an exact result,
obtained without approximations. We have computed numerically the Lyapunov
exponent by measuring $\overline{\log E}$ in the time-asymp\-to\-tic regime
and thus confirmed the validity of (\ref{Lyapunovapp}).

\end{document}